\documentclass[10pt]{article}
\begin{document}

\title{Generating solutions for charged stellar models in general relativity}
\author{B.V.Ivanov \\
%EndAName
Institute for Nuclear Research and Nuclear Energy, \\
Bulgarian Academy of Science, \\
Tzarigradsko Shausse 72, Sofia 1784, Bulgaria}
\maketitle

\begin{abstract}
It is shown that the expressions for the tangential pressure, the anisotropy
factor and the radial pressure in the Einstein - Maxwell equations may serve
as generating functions for charged stellar models. The latter can
incorporate an equation of state when the expression for the energy density
is also used. Other generating functions are based on the condition for the
existence of conformal motion (conformal flatness in particular) and the
Karmarkar condition for embedding class one metrics, which do not depend on
charge. In all these cases the equations are linear first order differential
equations for one of the metric components and Riccati equations for the
other. The latter may be always transformed into second order homogenous
linear differential equations. These conclusions are illustrated by numerous
particular examples from the study of charged stellar models.
\end{abstract}

\section{Introduction}

Gravitation is governed by the Einstein equations of general relativity in
the simplest case. The Einstein-Maxwell equations are a system of highly
non-linear differential second order equations in partial derivatives. In
astrophysics spherical symmetry is usually used, which reduces in the static
case the differential equations to ordinary ones and the derivatives are
with respect to the radius. The metric is diagonal with just two components.
In canonical comoving coordinates there are three Einstein equations for six
unknowns - the two metric potentials and the four components of the
energy-momentum tensor $T_{ab}$, namely, the energy density $\mu $, the
radial and the tangential pressures $p_r$ and $p_t$ and the charge $l$. Thus
the fluid is anisotropic, which is backed by arguments for compact objects
with very high density \cite{Ruder} and by a number of other reasons \cite
{HS}, \cite{BIJTP}.

On one side these equations present expressions for the components of the
energy-momentum tensor. On the other side the metric potentials enter in a
rather involved way as they are obtained from the Ricci tensor and scalar.
The equations remain non-linear for the metric. Durgapal and Banerjee \cite
{DB} showed that in the perfect fluid case the Einstein equations are linear
of first order for a function of $g_{11}$ and the equations for $p_t$ and
the anisotropy factor $\Delta $ are linear of second order for a function of 
$g_{00}$. Later, these findings were generalised for charged anisotropic
fluid. The reason for this simplification was partly clarified in \cite{Bone}
and is due to the fact that the Einstein equation for $p_t$ is a Riccati
equation. It was also shown there that the Einstein equations may serve as
generating functions for stellar model solutions, similar to the case of $%
\Delta $ \cite{nd}. The existence of an EOS leads to a relation between the
metric potentials.

Something more, there are common features between the generating functions
based on the equations for $p_t$ and $\Delta $ and other ways to generate a
solution, like conformal flatness, conformal motion or the possibility to
embed the spacetime in a flat 5-dimensional spacetime, namely they are also
linear or Riccati, which in the last case is truncated to a Bernoulli
equation.

In the present paper we discuss the charged anisotropic case in a systematic
way. We shall not study the many conditions for physical viability of the
solution, but concentrate on the mathematical issues and back them with
plenty of concrete examples from the literature in the corresponding section.

In Sect. 2 the Einstein - Maxwell equations are given, as well as some
characteristics of the model and the equations for the anisotropy factor,
the existence of conformal motion or flatness in particular, and the
Karmarkar condition. In Sect. 3 we give the three types of differential
equations that are used in the following and list some of their properties,
using the notation of handbooks on solutions of ordinary differential
equations. In Sect. 4 a generating function, based on the expression for the
radial pressure is discussed. When an EOS is imposed, the expression for the
energy density is also necessary. Sect. 5 gives generating function based on
the expressions for the tangential pressure. The well-known generating
function, based on the anisotropy factor, is generalised to the charged
case. In Sect. 6 we discuss the metric potentials as generating functions,
with or without a relation between them. Sect. 7 deals with generating
solutions when the charge is not given beforehand. Sect. 8 provides some
discussion.

\section{Einstein-Maxwell equations and definitions}

The interior of static spherically symmetric stars is described by the
canonical line element 
\begin{equation}
ds^2=e^\nu dt^2-e^\lambda dr^2-r^2\left( d\theta ^2+\sin ^2\theta d\varphi
^2\right) ,  \label{one}
\end{equation}
where $\lambda $ and $\nu $ depend only on the radial coordinate $r$. The
energy-momentum tensor reads 
\begin{equation}
T_{\alpha \beta }=\left( \mu +p_t\right) u_\alpha u_\beta +p_tg_{\alpha
\beta }+\left( p_r-p_t\right) \chi _\alpha \chi _\beta +E_{\alpha \beta }.
\label{two}
\end{equation}

Here $\mu $ is the energy density, $p_r$ is the radial pressure, $p_t$ is
the tangential pressure, $u^\alpha $ is the four-velocity of the fluid, $%
\chi ^\alpha $ is a unit spacelike vector along the radial direction and $%
E_{\alpha \beta }$ is the electromagnetic energy tensor.

We have 
\begin{equation}
E_{\alpha \beta }=\frac 1{4\pi }\left( F_\alpha ^{\;\gamma }F_{\beta \gamma
}-\frac 14g_{\alpha \beta }F^{\gamma \delta }F_{\gamma \delta }\right) ,
\label{three}
\end{equation}
where $F_{\alpha \beta }$ is the electromagnetic field tensor. Its only
non-trivial component $F_{01}=-F_{10}=-\Phi ^{\prime }$ is expressed through
the four-potential, which has only a time component $\Phi $. The prime
stands for a radial derivative. The Maxwell equations yield 
\begin{equation}
\Phi ^{\prime }=\frac{e^{\nu /2+\lambda /2}l}{r^2},\quad l\left( r\right)
=4\pi \int_0^r\sigma e^{\lambda /2}r^2dr,  \label{four}
\end{equation}
where $\sigma $ is the charge density and $l\left( r\right) $ is the total
charge up to radius $r$. We use relativistic units with $G=1,c=1,k=8\pi $.

The Einstein equations read 
\begin{equation}
8\pi \mu +\frac{l^2}{r^4}=\frac 1{r^2}-\left( \frac 1{r^2}-\frac{\lambda
^{\prime }}r\right) e^{-\lambda },  \label{five}
\end{equation}

\begin{equation}
8\pi p_r-\frac{l^2}{r^4}=-\frac 1{r^2}\left( 1-e^{-\lambda }\right) +\frac{%
\nu ^{\prime }}re^{-\lambda },  \label{six}
\end{equation}
\begin{equation}
kp_t+\frac{l^2}{r^4}=\frac{e^{-\lambda }}4\left( 2\nu ^{\prime \prime }+\nu
^{\prime 2}+\frac{2\nu ^{\prime }}r-\nu ^{\prime }\lambda ^{\prime }-\frac{%
2\lambda ^{\prime }}r\right) ,  \label{seven}
\end{equation}
where $\mu $ is the matter density, $p_r$ is the radial pressure and $p_t$
is the tangential one.

The gravitational mass in a sphere of radius $r$ is given by 
\begin{equation}
\frac{2m}r=1-e^{-\lambda }+\frac{l^2}{r^2}.  \label{eight}
\end{equation}
which may be written also as 
\begin{equation}
e^{-\lambda }=1-\frac{2m}r+\frac{l^2}{r^2}.  \label{nine}
\end{equation}

The field equations do not contain $\nu $, but its first and second
derivative. It is related to the four-acceleration $a_1$, namely $2a_1=\nu
^{\prime }$.

As a whole, we have three field equations for six unknown functions: $%
\lambda ,\nu ,\mu ,p_r,$ $p_t$ and $l$. We can choose freely three of them,
but the model will be physically realistic if a number of regularity,
matching and stability conditions are satisfied too. Choosing $\lambda ,\nu
,l$ means to charge a neutral solution with the same $\lambda $ and $\nu $.
Then $p_r$ and $m$ increase, but $\mu $ and $p_t$ decrease.

Different constraints may be imposed on the system of Einstein - Maxwell
equations. One of them is the existence of an equation of state (EOS) $%
p_r=f\left( \mu \right) $.

Let us introduce the anisotropic factor $\Delta =p_t-p_r$. It measures the
anisotropy of the fluid. Eqs. (6, 7) give 
\begin{equation}
-8\pi \Delta -\frac{2l^2}{r^4}=e^{-\lambda }\left( -\frac{\nu ^{\prime
\prime }}2-\frac{\nu ^{\prime 2}}4+\frac{\nu ^{\prime }}{2r}+\frac
1{r^2}\right) +e^{-\lambda }\frac{\lambda ^{\prime }}2\left( \frac{\nu
^{\prime }}2+\frac 1r\right) -\frac 1{r^2}.  \label{ten}
\end{equation}
When $\Delta =0$ the fluid becomes perfect and all pressures are equal.
Charging a neutral solution decreases its $\Delta $.

The following two requirements may be imposed on the spacetime.

The first is conformally flat spacetime. It takes place when its Weyl tensor
vanishes. This is a particular case of spacetimes with conformal motion when
a Killing vector $\mathbf{K}$ exists. Then the following equation has to be
satisfied 
\begin{equation}
L_{\mathbf{K}}g_{ab}=2\psi g_{ab},  \label{eleven}
\end{equation}
where $L_{\mathbf{K}}$ is the Lie derivative operator and $\psi \left(
t,r\right) $ is the conformal factor. This implies the equation \cite{cm} 
\begin{equation}
2\nu ^{\prime \prime }+\nu ^{\prime 2}=\nu ^{\prime }\lambda ^{\prime }+%
\frac{2\nu ^{\prime }}r-\frac{2\lambda ^{\prime }}r+\frac 4{r^2}\left(
1+s\right) e^\lambda -\frac 4{r^2},  \label{twelve}
\end{equation}
where $s$ is a constant of integration. The spacetime is conformally flat
when $s=0$.

In recent years spacetimes, which are embeddings of class one, have been
widely discussed. They can be embedded in a 5-dimensional flat spacetime.
This requires the Karmarkar relation between the components of the Riemann
tensor \cite{Karmarkar} 
\begin{equation}
R_{1010}R_{2323}-R_{1212}R_{3030}=R_{1220}R_{1330}.  \label{thirteen}
\end{equation}
It transforms into a differential equation for $\lambda $ and $\nu $: 
\begin{equation}
2\frac{\nu ^{\prime \prime }}{\nu ^{\prime }}+\nu ^{\prime }=\frac{\lambda
^{\prime }e^\lambda }{e^\lambda -1}.  \label{fourteen}
\end{equation}

The charge does not enter Eqs (12, 14), hence, the system (5-7) represents
in these cases the charging of a neutral solution with conformal motion or
an embedding of class one.

\section{Types of equations}

We have shown in the uncharged case \cite{Bone} that Eqs (5, 6, 7, 10, 12)
are linear with respect to $y=e^{-\lambda }$ while Eq (14) is linear for $%
y=e^\lambda $. Eq (5) does not contain $a_1$, while Eq (6) gives an
expression for it. The others belong to three types of equations with
respect to $a_1$. They may be transformed into linear equations for $%
u=e^{\nu /2}$. Now we shall show that charging of the fluid does not alter
this properties. Charged anisotropic fluid is the general type of fluid in
the static case. All other characteristics like shear, expansion, two types
of viscosity, two types of radiation depend on time and vanish for static
solutions \cite{BVIone}.

To make the paper self-contained we give some properties of the types of
equations, which appear in the following. We stick to the standard
mathematical notation of handbooks on differential equations \cite{PZ}.
Except for the constants $C$ and $n$, the other letters designate functions
of $r$.

1) Linear equation. It reads 
\begin{equation}
gy^{\prime }=f_1y+f_0.  \label{fifteen}
\end{equation}
It is integrable and its general solution is 
\begin{equation}
y=Ce^F+e^F\int e^{-F}\frac{f_0}gdr,\quad F=\int \frac{f_1}gdr.
\label{sixteen}
\end{equation}

2) Bernoulli equation. It reads 
\begin{equation}
gy^{\prime }=f_1y+f_ny^n  \label{seventeen}
\end{equation}
and becomes a linear equation for $w=y^{1-n}$%
\begin{equation}
gw^{\prime }=\left( 1-n\right) f_1w+\left( 1-n\right) f_n,  \label{eighteen}
\end{equation}
so it is soluble in general. Using (16) its solution becomes 
\begin{equation}
y^{1-n}=Ce^F+\left( 1-n\right) e^F\int e^{-F}\frac{f_n}gdr,\quad F=\left(
1-n\right) \int \frac{f_1}gdr.  \label{nineteen}
\end{equation}

3) Riccati equation. It is given by 
\begin{equation}
gy^{\prime }=f_2y^2+f_1y+f_0  \label{tw}
\end{equation}
and no general solution is known. In particular cases it reduces to
integrable equations. Thus when $f_2=0$ it turns into a linear equation.
When $f_0=0$ it becomes a Bernoulli equation with $n=2$, so that $1/y$
satisfies a linear equation. Every Riccati equation may be transformed into
a canonical form with $f_1=0$. There is a general solution for the Riccati
equation when one particular solution $y_0$ is known: 
\begin{equation}
y=y_0+F\left( C-\int F\frac{f_2}gdr\right) ^{-1},\quad F=\exp \int \left(
2f_2y_0+f_1\right) \frac{dr}g.  \label{twone}
\end{equation}
This formula simplifies when two or more particular solutions are known.

This equation may be transformed into a second-order homogenous linear
equation for $u$ when the following substitution is made 
\begin{equation}
u=\exp \left( -\int \frac{f_2}gydr\right) .  \label{twtwo}
\end{equation}
Namely, Eq. (20) becomes 
\begin{equation}
g^2f_2u^{\prime \prime }+g\left[ f_2g^{\prime }-gf_2^{\prime }-f_1f_2\right]
u^{\prime }+f_0f_2^2u=0.  \label{twthree}
\end{equation}
In the case $f_2=-g$ Eqs. (22, 23) simplify considerably 
\begin{equation}
u=\exp \int ydr,  \label{twfour}
\end{equation}
\begin{equation}
f_2u^{\prime \prime }+f_1u^{\prime }+f_0u=0.  \label{twfive}
\end{equation}
The substitution $y=u^{\prime }/u$ leads back to Eq (20).

\section{The energy density and the radial pressure}

In the following we consider $l$ as known. Eq (5) for the energy density
does not contain $a_1$. It is linear with respect to $y=e^{-\lambda }$ and
can be written as 
\begin{equation}
ry^{\prime }=-y+1-8\pi \mu r^2-\frac{l^2}{r^2}.  \label{twsix}
\end{equation}
Eq (9) may be written as 
\begin{equation}
y=1-\frac{2m}r+\frac{l^2}{r^2}.  \label{twseven}
\end{equation}
Any equation, linear in $y$ may be transformed into an equation, linear in $%
m $ with the use of the above formula.

Eq (6) for the radial pressure may be written as 
\begin{equation}
8\pi p_rr^2=\left( 2a_1r+1\right) y-1+\frac{l^2}{r^2}.  \label{tweight}
\end{equation}
It may be regarded as an expression for $p_r$ or $y$%
\begin{equation}
y=\frac{8\pi p_rr^2+1-\frac{l^2}{r^2}}{2ra_1+1},  \label{twnine}
\end{equation}
or $a_1$%
\begin{equation}
2a_1=\nu ^{\prime }=\frac{8\pi p_rr^2+1-y-\frac{l^2}{r^2}}{ry}.
\label{thirty}
\end{equation}
The potential $\nu $ is found by a simple quadrature.

Thus, Eq (28), which contains $p_r$, $y$ and $a_1$, is the simplest
generating function for any of them, when the other two are known. Solutions
with given $y$ (or $m$) and $p_r$ may be found in \cite{lpone}, \cite{lptwo}%
, \cite{lpthree}, \cite{lpfour}, \cite{lpfive}.

An EOS can be incorporated in this scheme, $p_r=f\left( \mu \right) $ or 
\begin{equation}
2rya_1=1-\frac{l^2}{r^2}-y+8\pi r^2f\left( -\frac{ry^{\prime }+y-1+\frac{l^2%
}{r^2}}{8\pi r^2}\right) ,  \label{thone}
\end{equation}
which follows from Eqs (26, 28). Obviously, the resulting equation is not
linear in $y$ in general, but still may be solvable by choosing an ansatz
for $y$. Anyway, its an expression for $a_1$ in terms of $y$ and is a
relation between the metric potentials. Mainly EOS with ansatz for $y$ were
used. Thus quadratic EOS is discussed in \cite{QEOSone}, \cite{QEOStwo},
polytropic EOS in \cite{pone}, \cite{ptwo}, \cite{pthree}, other EOS in \cite
{EOSone}, \cite{EOStwo}, \cite{EOSthree}.

A special case is the linear EOS (LEOS) $p_r=a\mu -b$ with constant $0\leq
a\leq 1$ and the bag constant $b\geq 0$, which includes also the case $p_r=0$%
. Eq (31) becomes 
\begin{equation}
2rya_1=\left( a+1\right) \left( 1-\frac{l^2}{r^2}-y\right) -ary^{\prime
}-8\pi br^2.  \label{thtwo}
\end{equation}
This is an expression for $a_1$ when $y$ and $q$ are given and was used in 
\cite{LEOSone}, \cite{LEOStwo}, \cite{LEOSthree}, \cite{LEOSfour}, \cite
{LEOSfive}, \cite{LEOSsix}, \cite{LEOSseven}, \cite{LEOSeight}, \cite
{LEOSnine}.

Eq (32) is also a linear equation for $y$ 
\begin{equation}
ary^{\prime }=-\left( 2ra_1+a+1\right) y+\left( a+1\right) \left( 1-\frac{l^2%
}{r^2}\right) -8\pi br^2.  \label{ththree}
\end{equation}
It can be solved by Eq (16) when $a,b,a_1$ are known. The factor $F$ is the
same as in the uncharged case (cite). Eq (16) gives a singular $e^F$ for $%
r=0 $, hence $C=0$. Then the solution is 
\begin{equation}
y=\frac{\int \left[ \left( a+1\right) \left( 1-\frac{l^2}{r^2}\right) -8\pi
br^2\right] \left( re^\nu \right) ^{1/a}dr}{a\left( r^{a+1}e^\nu \right)
^{1/a}}.  \label{thfour}
\end{equation}

The relation between the energy density and the mass is more complicated for
a charged fluid. Integrating Eq (5) and using formula (9) we get 
\begin{equation}
m=\frac 12\int \left( 8\pi \mu r^2+\frac{l^2}{r^2}\right) dr+\frac{l^2}{2r}.
\label{thfive}
\end{equation}
This expression reduces to the one in the neutral case when $l=0$. It may be
written also as 
\begin{equation}
m^{\prime }=4\pi \mu r^2+\frac{ll^{\prime }}r.  \label{thsix}
\end{equation}
This formula shows that when we pass from $y$ to $m$ Eq (26) simplifies.

\section{The tangential pressure and the anisotropic factor}

Eq (7) is an expression for $p_t$ and can be written as a linear equation
for $y$ 
\begin{equation}
\frac 12\left( a_1+\frac 1r\right) y^{\prime }=-\left( a_1^{\prime }+a_1^2+%
\frac{a_1}r\right) y+8\pi p_t+\frac{l^2}{r^4}.  \label{thseven}
\end{equation}
Its solution from Eq (16) reads 
\begin{equation}
y=e^F\left( C+\int ze^\nu e^{2\int \frac{dr}{r^2z}}\left( 16\pi p_t+\frac{l^2%
}{r^4}\right) dr\right) ,  \label{theight}
\end{equation}
where 
\begin{equation}
a_1+\frac 1r=\frac{\nu ^{\prime }}2+\frac 1r\equiv z,  \label{thnine}
\end{equation}
\begin{equation}
e^F=z^{-2}e^{-\nu }e^{-2\int \frac{dr}{r^2z}}.  \label{forty}
\end{equation}
The term $e^F$ is the same as in the uncharged case. Due to Eq (9), Eq (37)
is also linear with respect to the mass.

Eq (7) is also a Riccati equation for $a_1$%
\begin{equation}
ya_1^{\prime }=-ya_1^2-\left( \frac yr+\frac{y^{\prime }}2\right) a_1-\frac{%
y^{\prime }}{2r}+8\pi p_t+\frac{l^2}{r^4}  \label{foone}
\end{equation}
and may be solved for particular choices of $y$ and $p_t$. It can be
transformed into a linear second order homogenous differential equation
following Eqs (24, 25) 
\begin{equation}
yu^{\prime \prime }+\left( \frac yr+\frac{y^{\prime }}2\right) u^{\prime
}+\left( \frac{y^{\prime }}{2r}-8\pi p_t-\frac{l^2}{r^4}\right) u=0,
\label{fotwo}
\end{equation}
where 
\begin{equation}
u=e^{\nu /2}  \label{fothree}
\end{equation}
Sometimes it may be solved easier than the original Riccati equation, since
many special functions are defined by such equations. It remains in the same
time linear (and integrable) first order equation for $y=e^{-\lambda }$ or $%
m $. It can be called a double linear equation. Thus, like $p_r$, the
expression (7) for $p_t$ is a generating function for charged stellar
models, when two of the quantities $p_r$, $y$ (or $m$) and $a_1$ are known.

The generating functions based on $\Delta $ are found in a similar way. Eq
(10) is linear with respect to $y$ (or $m$) and may be rewritten as 
\begin{equation}
\left( a_1+\frac 1r\right) y^{\prime }=-2\left( a_1^{\prime }+a_1^2-\frac{a_1%
}r-\frac 1{r^2}\right) y-2\left( \frac 1{r^2}-8\pi \Delta -\frac{2l^2}{r^4}%
\right) .  \label{fofour}
\end{equation}
After some transformations it becomes 
\begin{equation}
y^{\prime }=-2\left( \frac{z^{\prime }}z+z-\frac 3r+\frac 2{r^2z}\right)
y-\frac 2z\left( \frac 1{r^2}-8\pi \Delta -\frac{2l^2}{r^4}\right) .
\label{fofive}
\end{equation}
This is the generalisation of Eq (8) from \cite{nd} to the charged case when
the different definition of their $\Delta $ is taken into account and is
still integrable. The result is 
\begin{equation}
y=r^6z^{-2}e^{-\int \left( \frac 4{r^2z}+2z\right) dr}\left[ C-2\int
r^{-8}z\left( 1-8\pi \Delta r^2-\frac{2l^2}{r^2}\right) e^{\int \left( \frac
4{r^2z}+2z\right) dr}dr\right] .  \label{fosix}
\end{equation}
The generating potentials are $\Delta $, $z$ and $l$, the second, due to Eq
(39), is equivalent to $a_1$. This generating function encompasses the
important cases of charged perfect fluid when $\Delta =0$ \cite{Btwo} and
neutral perfect fluid when $\Delta =0$, $l=0$. Solutions with given $\Delta
,a_1$ and $q$ are discussed in \cite{ndone}, \cite{ndtwo}, \cite{ndthree},
where the mass is used instead of $y$, \cite{ndfour}, \cite{ndfive}, \cite
{ndsix}, \cite{ndseven}, \cite{ndeight}, \cite{ndnine}, \cite{ndten}. There
are also solutions with $\Delta =0$ \cite{ndeleven}, \cite{ndtwelve}.

Eq (44) is also a Riccati one for $a_1$, the Riccati structure $a_1^{\prime
}+a_1^2$ being brought in $\Delta $ by $p_t$. It can be written as 
\begin{equation}
2ya_1^{\prime }=-2ya_1^2+\left( \frac{2y}r-y^{\prime }\right) a_1+\frac{%
2y-2-ry^{\prime }}{r^2}+16\pi \Delta +\frac{4l^2}{r^4}  \label{foseven}
\end{equation}
and solved for particular $\Delta $, $y$ and $l$. Finally, it can be
linearised following Eqs (24, 25) into 
\begin{equation}
-2yu^{\prime \prime }+\left( \frac{2y}r-y^{\prime }\right) u^{\prime
}+\left( \frac{2y-2-ry^{\prime }}{r^2}+16\pi \Delta +\frac{4l^2}{r^4}\right)
u=0,  \label{foeight}
\end{equation}
where $u$ is given by Eq (43). Thus, once again, Eq (48) is doubly linear,
like Eq (42). Solutions of this equation were presented in \cite{ldone}, 
\cite{ldtwo}, \cite{ldthree}, \cite{ldfour} and with $\Delta =0$ in \cite
{ldfive}. In total, Eq (10) is a generating function for stellar models,
when $l$ and two of the quantities $\Delta $, $y$ (or $m$) and $a_1$ are
known. The differential equations for $y$ and $u$ are linear.

\section{The metric potentials as generating functions}

The simplest way to generate solutions in the charged case is to choose
independently the two generating potentials $\lambda $ and $\nu $ and add to
them a third potential $l$. Thus any neutral solution may be charged \cite
{lnone}, \cite{lntwo}.

Some important stellar models require a relation between $\lambda $ and $\nu 
$, reducing the generating functions to two. For example this is the case of
charged perfect fluid, when in Eq. (10) $\Delta =0$. Similar example are
spacetimes admitting conformal motion. The metric potentials of such
spacetimes satisfy Eq (12). In \cite{cm} this equation is solved by a series
of transformations. Surprisingly, it is also a linear equation in $y$ (or $m$%
) and a Riccati equation for $a_1$. It can be written as \cite{Bone} 
\begin{equation}
\left( \frac 1r-a_1\right) y^{\prime }=2\left( a_1^{\prime }+a_1^2-\frac{a_1}%
r+\frac 1{r^2}\right) y-\frac{2\left( 1+s\right) }{r^2}  \label{fonine}
\end{equation}
or 
\begin{equation}
2ya_1^{\prime }=-2ya_1^2+\left( \frac{2y}r-y^{\prime }\right) a_1+\frac{%
y^{\prime }}r+\frac{2\left( 1+s\right) -2y}{r^2}.  \label{fifty}
\end{equation}
Once again $g=-f_2$ in Eq (20), so it may be transformed into a linear
equation, analogous to Eq (25) 
\begin{equation}
-2yu^{\prime \prime }+\left( \frac{2y}r-y^{\prime }\right) u^{\prime
}+\left( \frac{y^{\prime }}r+\frac{2\left( 1+k\right) -2y}{r^2}\right) u=0,
\label{fione}
\end{equation}
where $u$ is given by Eq (24). In \cite{cm} its solution was found and
possesses three branches 
\begin{equation}
e^\nu =Ar\exp \left( \sqrt{1+s}\int \frac{e^\lambda }rdr\right) +Br\exp
\left( -\sqrt{1+s}\int \frac{e^\lambda }rdr\right) ,\quad 1+s>0,
\label{fitwo}
\end{equation}
\begin{equation}
e^\nu =Ar\int \frac{e^\lambda }r+Br,\quad 1+s=0,  \label{fithree}
\end{equation}
\begin{equation}
e^\nu =Ar\exp \left( \sqrt{-\left( 1+s\right) }\int \frac{e^\lambda }%
rdr\right) +Br\exp \left( -\sqrt{-\left( 1+s\right) }\int \frac{e^\lambda }%
rdr\right) ,\quad 1+s<0.  \label{fifour}
\end{equation}
and do not depend on the charge. Solutions with conformal motion were
discussed recently \cite{cmone}, \cite{cmtwo}, \cite{cmthree}. In \cite
{cmfour} these expressions were put into Eq (44) and another equation for $y$
arises, which is simpler. Solutions based on $\psi $ in Eq (12) were studied
in \cite{cmfive}.

Another example is the Karmarkar condition for embedding of class one, Eq
(14). It was discussed in \cite{Bone}. It may be written as 
\begin{equation}
a_1^{\prime }=-a_1^2+\left[ \ln \left( \frac{1-y}y\right) \right] ^{\prime }%
\frac{a_1}2.  \label{fifive}
\end{equation}
The would be Riccati equation becomes a Bernoulli one (see Eq (17)) with $n=2
$. It is also a Bernoulli equation for $y$ 
\begin{equation}
-\frac{a_1}2y^{\prime }=\left( a_1^{\prime }+a_1^2\right) y-\left(
a_1^{\prime }+a_1^2\right) y^2.  \label{fisix}
\end{equation}
All these equations are solvable. Their integration may be done directly,
without using the general formulas and we obtain the well-known results 
\begin{equation}
e^\lambda =C\nu ^{\prime 2}e^\nu +1,  \label{fiseven}
\end{equation}
\begin{equation}
e^\nu =\left( A+B\int \sqrt{e^\lambda -1}dr\right) ^2.  \label{fieight}
\end{equation}
where $A,B,C$ are integration constants. Thus when one of the metric
coefficients is given, we can find the other. The solution may be charged by
introducing a known $q$. It only changes the system of Einstein-Maxwell
equations (5-7). Solutions with given $\lambda $ and $q$ were found in \cite
{kone}, \cite{ktwo}, \cite{ktwoa}. Solutions with known $\nu $ and $q$ were
studied in \cite{kthree}, \cite{kfour}.

\section{Solutions when the charge is not given beforehand}

Up to now we have discussed cases with given $l^2$. However, solutions may
be found when this is not so. It is clear that the LEOS Eq (33) is also an
expression for $l^2$%
\begin{equation}
\left( a+1\right) \frac{l^2}{r^2}=-ary^{\prime }-\left( 2ra_1+a+1\right)
y+a+1-8\pi br^2.  \label{finine}
\end{equation}
We can find $l^2$ when $y$ and $a_1$ are known, i.e. when $\lambda $ and $%
\nu $ are known and a LEOS is given \cite{qone}.

Eq (44) can also serve as an expression for $l^2$%
\begin{equation}
\frac{4l^2}{r^4}=\left( a_1+\frac 1r\right) y^{\prime }+2\left( a_1^{\prime
}+a_1^2-\frac{a_1}r-\frac 1{r^2}\right) y+2\left( \frac 1{r^2}-8\pi \Delta
\right)  \label{sixty}
\end{equation}
when $y$, $a_1$ and $\Delta $ are known. Thus, we can give an ansatz for $%
\lambda $, add the Karmarkar condition to find $\nu $ and set $\Delta =0$ 
\cite{qtwo}, \cite{qtwoa}, \cite{qtwob}, \cite{qtwoc}. The isotropic
condition may be written also as Eq (48), another expression for $l^2$ 
\begin{equation}
-\frac{4l^2}{r^4}=-2y\frac{u^{\prime \prime }}u+\left( \frac{2y}r-y^{\prime
}\right) \frac{u^{\prime }}u+\frac{2y-2-ry^{\prime }}{r^2}+16\pi \Delta ,
\label{sione}
\end{equation}
where $u=e^{\nu /2}$. Fixing $y$, setting $\Delta =0$ (perfect fluid) and
with some simplifying assumption one can solve this equation \cite{qtwod}.

Together, Eqs (59, 60) give another linear equation for $y$ which depends
only on $a_1$ and $\Delta $. Solving it, we find $y$ and then $l^2$ from any
of Eqs (59, 60). This approach was used in \cite{qthree}, \cite{qfour}.

One can use another EOS, e.g. the Chaplygin EOS 
\begin{equation}
p_r=\alpha _1\mu -\frac{\alpha _2}\mu ,  \label{sitwo}
\end{equation}
where $\alpha _1,\alpha _2$ are positive constants. Summing Eqs (5, 6) we
obtain 
\begin{equation}
p_r=G\left( \lambda ,\nu \right) -\mu ,  \label{sithree}
\end{equation}
where $G$ is some function. Replacing (63) into (62) yields a quadratic
equation for $\mu $, which is solvable 
\begin{equation}
\left( \alpha _1+1\right) \mu ^2-G\mu -\alpha _2=0.  \label{sifour}
\end{equation}
The metric components $\lambda $ and $\nu $ may be supplied directly \cite
{qfive}. Another way is to fix one of them, e.g. $\nu $ and impose the
Karmarkar condition to find $\lambda $ \cite{qsix}.

Similar is the situation with the quadratic EOS 
\begin{equation}
p_r=\alpha _1\mu ^2+\alpha _2\mu +\alpha _3.  \label{sifive}
\end{equation}
Eq (63) shows that this is a quadratic equation for $\mu $%
\begin{equation}
\alpha _1\mu ^2+\left( \alpha _2+1\right) \mu +\alpha _3-G=0  \label{sisix}
\end{equation}
and may be solved too.

Another popular EOS is the modified Van der Vaals one 
\begin{equation}
p_r=\alpha _1\mu ^2+\frac{\alpha _2\mu }{1+\alpha _3\mu }.  \label{siseven}
\end{equation}
It becomes 
\begin{equation}
\alpha _1\alpha _3\mu ^3+\left( \alpha _1+\alpha _3\right) \mu ^2+\left(
\alpha _2+1-\alpha _3G\right) \mu -G=0.  \label{sieight}
\end{equation}
This is a cubic equation for $\mu $ and is still solvable.

Finally, let us discuss the polytropic EOS 
\begin{equation}
p_r=\alpha \mu ^{1+\frac 1N},  \label{sinine}
\end{equation}
where $\alpha $ is a constant and $N$ is the polytropic index. It can be
written with the help of Eq (63) as 
\begin{equation}
\alpha ^N\mu ^{N+1}-\left( G-\mu \right) ^N=0.  \label{seventy}
\end{equation}
This equation is quadratic for $N=1$, cubic for $N=2$ and quartic for $N=3$
and therefore solvable for $\mu $ for these values of $N$.

\section{Discussion}

In a previous paper \cite{Bone} we have studied the existence of generating
functions, giving solutions for uncharged stellar models. In the present one
we do the same for charged models. The addition of charge does not alter the
general scheme of using the Einstein equations as generating functions. Now
three of the four characteristics of the model should be given - $%
y=e^{-\lambda },a_1=\nu ^{\prime }/2,l$ and either $p_r$, $p_t$ or $\Delta $%
. This approach is greatly simplified, because the Einstein equations with
charge are still linear first order differential equations for $y$ and
Riccati equations for the four-acceleration $a_1$. The first are always
integrable in quadratures, while the second are integrable in many
particular cases. There is a standard mathematical procedure to transform
them into linear homogenous differential equations of second order for $%
u=e^{\nu /2}$. They are the ''missing link'' between the original form of
the Einstein equations and their linear version, which appears out of
nowhere in \cite{DB} and holds also for anisotropic and charged fluids. The
source of the Riccati structure $a_1^{\prime }+a_1^2$ still comes from the
component $R_{0101}$ of the Riemann tensor, whose expression is the same in
the charged case. Eq (8) shows that the mass $m$ still satisfies a linear
equation and may replace $y$.

There are two main ways of generating solutions. The first one accepts that $%
l$ is a given function. Then the stellar models for neutral fluids are just
charged. The simplest generating function is Eq (6) for the $p_r$, which is
an expression for $p_r$, $a_1$, $y$ or $l^2$ without solving any equations.
Eq (5) for the energy density cannot be used as a generating function,
because it does not contain $\nu $. However, when the model has an EOS, the
combination of Eqs (5) and (6) works as a generating function, producing a
relation between the two metric potentials and $l^2$. The equation for $%
\Delta $ was used in \cite{nd} to obtain $\lambda $ when $\nu $ and $\Delta $
are given. It becomes a generating function for perfect fluid models when $%
\Delta =0$. We have generalised it to the charged case. Eq (7) for $p_t$ can
play a similar role.

Of course, the simplest generating potentials are $\lambda $ and $\nu $ and $%
l$. There are physical reasons that sometimes impose a relation between the
metric components. This happens when an EOS exists. It depends on $l^2$.

A second important case is that of spacetimes admitting conformal motion
(conformal flatness in particular). The surprising fact is that this
relation is also a linear differential equation for $y$ or $u$ and a Riccati
one for $a_1$. It does not depend on the charge.

A third well-known example are spacetimes of embedding class one, obeying
the Karmarkar condition. Here there is a minor difference - the relation is
a linear equation for $1/y$ and a Bernoulli equation for $y$, which is also
integrable. Furthermore, it is a Bernoulli equation with quadratic term for $%
a_1$. The Riccati structure, discussed above, is still present but there is
no free term. It also is charge independent.

The second way of generating solutions is when $l$ is not given beforehand.
In the previous section we have outlined the different ways to solve the
Einstein equations in this case. One of them relies on the existence of an
EOS. It leads to algebraic equations for most of the popular EOS, which are
soluble up to fourth order included.

It is interesting whether in some of the alternative theories of gravitation
similar simplifications occur.

\end{document}